\documentclass[12pt,reqno]{amsart}
\usepackage{amsmath}
\usepackage{amsfonts}
\usepackage{amsthm}
\usepackage{graphicx}

\newenvironment{remark}{\medskip\noindent{\it Remark:\/} }{\medskip}
\numberwithin{equation}{section}

\def\Ai{\mathop{\mathrm{Ai}}\nolimits}
\def\diag{\mathop{\mathrm{diag}}}

\newcommand{\R}{{\mathbb R}}

\renewcommand{\Re}{{\operatorname{Re\,}}}
\renewcommand{\Im}{{\operatorname{Im\,}}}

\newcommand{\Tr}{{{\operatorname{Tr}}}}

\newcommand{\Ga}{\Gamma}

\newcommand{\la}{\lambda}

\newcommand{\Om}{\Omega}

\newtheorem{theo}{{\sc \bf Theorem}}[section]

\DeclareMathOperator{\im}{Im}

\begin{document}
\title{Large $n$ limit of Gaussian random matrices with external
source, part II}
\author{Alexander I. Aptekarev}
\address{Keldysh Institute of Applied Mathematics, Russian Academy of Sciences,
Miusskaya Square 4, Moscow 125047, Russia}
\email{aptekaa@keldysh.ru}
\author{Pavel M. Bleher}
\address{Department of Mathematical Sciences,
Indiana University-Purdue University Indianapolis,
402 N. Blackford St., Indianapolis, IN 46202, U.S.A.}
\email{bleher@math.iupui.edu}
\author{Arno B.J. Kuijlaars}
\address{Department of Mathematics,
Katholieke Universiteit Leuven, Celestijnenlaan 200 B, B-3001
Leuven, Belgium}
\email{arno@wis.kuleuven.ac.be}

\thanks{The first and third author are supported in part
by INTAS Research Network NeCCA 03-51-6637 and by NATO Collaborative Linkage
Grant PST.CLG.979738. The first author is supported in part
by RFBR 02-01-00564 and the program ``Modern problems of theoretical mathematics'' RAS(DMS).
The second author is supported in part
by the National Science Foundation (NSF) Grant DMS-0354962.
The third author is supported in part by FWO-Flanders projects G.0176.02 and G.0455.04 and by
K.U.Leuven research grant IOT/04/24.}

\date{\today}

\begin{abstract}
We continue the study of the Hermitian random matrix ensemble
with external source
\[ \frac{1}{Z_n} e^{-n \Tr(\frac{1}{2}M^2 -AM)} dM \]
where $A$ has two distinct eigenvalues $\pm a$ of equal multiplicity.
This model exhibits a phase transition for the value $a=1$, since
the eigenvalues of $M$ accumulate on two intervals for $a > 1$,
and on one interval for $0 < a < 1$.
The case $a > 1$ was treated in part I, where it was proved that
local eigenvalue correlations have the universal limiting behavior
which is known for unitarily invariant random matrices, that is,
limiting eigenvalue correlations are expressed in terms of
the sine kernel in the bulk of the spectrum, and in terms
of the Airy kernel at the edge.
In this paper we establish the same results for the case $0 < a < 1$.
As in part I we apply the Deift/Zhou steepest descent analysis
to a $3 \times 3$-matrix Riemann-Hilbert problem.
Due to the different structure of an underlying
Riemann surface, the analysis includes an additional step
involving a global opening of lenses, which is a new
phenomenon in the steepest descent analysis of Riemann-Hilbert
problems.
\end{abstract}

\maketitle

\section{Introduction}
This paper is a continuation of \cite{BK2} to which we will frequently
refer in this paper. It will be followed by a third part \cite{BK4}, which
deals with the critical case. In these papers, we study the random
matrix ensemble with external source $A$
\begin{equation}\label{extsource1}
    \mu_n(dM)=\frac{1}{Z_n}\,e^{-n\Tr(V(M)-AM)}dM,
\end{equation}
defined on $n \times n$ Hermitian matrices $M$, with Gaussian potential
\begin{equation} \label{gausspotential}
    V(M) = \frac{1}{2}M^2
    \end{equation}
and with external source
\begin{equation}\label{extsource2}
    A=\diag(\underbrace{a,\ldots,a}_{n/2},\underbrace{-a,\ldots,-a}_{n/2}).
\end{equation}
In the physics literature, the ensemble (\ref{extsource1}) was
studied in a series of papers of Br\'ezin and Hikami
\cite{BH1}-\cite{BH5}, and P. Zinn-Justin \cite{ZJ1}, \cite{ZJ2}.
Our aim is to obtain rigorous results on eigenvalue correlations
in the large $n$-limit using the steepest descent / stationary
phase method for Riemann-Hilbert (RH) problems \cite{DZ1}, thereby
extending the works \cite{BI1}, \cite{BI2}, \cite{DKMVZ1},
\cite{DKMVZ2}, \cite{KV1}, \cite{KV2} who treated the unitary invariant case (i.e.,
$A=0$) with RH techniques.

While the unitary invariant case is connected with orthogonal
polynomials \cite{Deift,Mehta}, the ensemble (\ref{extsource1}) is connected with
multiple orthogonal polynomials \cite{BK1}. These are characterized
by a matrix RH problem \cite{VAGK}, and the eigenvalue correlation
kernel of (\ref{extsource1}) has a direct expression in terms of
the solution of this RH problem, see \cite{BK1}, \cite{DK} and
also formula (\ref{kernelK}) below.
The RH problem for (\ref{extsource1}) has size $(r+1) \times (r+1)$
if $r$ is the number of distinct eigenvalues of $A$. So with
the choice (\ref{extsource2}), the RH problem is $3 \times 3$-matrix
valued.

The asymptotic analysis of RH problems has been mostly restricted
to the $2 \times 2$ case. The analysis of larger size RH problems presents
some novel technical features as already demonstrated in \cite{BK2} and
\cite{KVAW}. In the present paper another new feature appears,
namely at a critical stage in the analysis we perform a
{\em global opening of lenses}. This global opening of lenses requires
a global understanding of an associated Riemann surface,
which is explicitly known for the Gaussian case (\ref{gausspotential}).
This is why we restrict ourselves to (\ref{gausspotential}) although
in principle our methods are applicable to more general polynomial $V$.

The Gaussian case has some special relevance in its own right as well.
Indeed, first of all we note that for (\ref{gausspotential}) we can
complete the square in (\ref{extsource1}), and then it follows that
\begin{equation} \label{randomplusdet}
    M = M_0 + A
\end{equation}
where $M_0$ is a GUE matrix. So in the Gaussian case the ensemble
(\ref{extsource1}) is an example of a random + deterministic model,
see also \cite{BH1}-\cite{BH5}.

A second interpretation of the Gaussian model comes from non-intersecting
Brownian paths. This can be seen from the joint probability
density for the eigenvalues of $M$, which by the HarishChandra/Itzykson-Zuber formula
\cite{IZ}, \cite{Mehta}, takes the form
\begin{equation} \label{jointpdf}
    \frac{1}{\tilde{Z}_n}
    \prod_{1 \leq j < k \leq n} (\lambda_j-\lambda_k) \,
    \det\left( e^{n \lambda_j a_k} \right)_{j,k=1}^n
    \prod_{j=1}^n e^{-\frac{1}{2} n \lambda_j^2}
\end{equation}
for the case (\ref{gausspotential}). Here $a_1, \ldots, a_n$ are the eigenvalues of $A$,
which are assumed to be all distinct in (\ref{jointpdf}). In the case of coinciding
eigenvalues of $A$ we have to take the appropriate limit of (\ref{jointpdf}),
see formula (3.17) in \cite{BK1}.

Formula (\ref{jointpdf}) also arises as the distribution of
non-intersecting Brownian paths. Consider $n$ independent Brownian motions
(in fact Brownian bridges) on the line, starting at some fixed points
$s_1 < s_2 < \cdots < s_n$ at time $t=0$, ending at some fixed points
$b_1 < b_2 < \ldots < b_n$ at time $t=1$, and conditioned not to
intersect for  $t \in (0,1)$. Then by a theorem of Karlin and McGregor
\cite{KMG}, the joint probability density of the positions of the
Brownian bridges at time $t \in (0,1)$ is given by
\begin{equation}\label{b1}
p_n(x_1,\dots,x_n)=\frac{1}{C_n}\det(p(s_j,x_k;t))_{j,k=1}^n
\det(p(x_j,b_k;1-t))_{j,k=1}^n
\end{equation}
where $p(x,y;t)$ is the transition kernel of the Brownian motion
and $C_n$ is a normalization constant.
Let us consider a scaled Brownian motion for which
\begin{equation}\label{b3}
p(x,y;t)=\sqrt{\frac{n}{2\pi t}}\,e^{-\frac{n(x-y)^2}{2t}}
\end{equation}
and let us take a limit when all initial points $s_j$ converge
to the origin. In this case formula (\ref{b1}) takes the form
\begin{equation}\label{b5}
p_n(x_1,\dots,x_n)=\frac{1}{\bar C_n} \prod_{1 \leq j < k \leq n} (x_j-x_k)
\, \det\left(e^{\frac{n x_jb_k}{1-t}}\right)_{j,k=1}^n
\prod_{j=1}^n e^{-\frac{n}{2t(1-t)}x_j^2}.
\end{equation}
This coincides with (\ref{jointpdf}) if we make the identifications
\begin{equation}\label{b6}
\lambda_j = \frac{x_j}{\sqrt{t(1-t)}},
\qquad a_k =  b_k \sqrt{\frac{t}{1-t}}.
\end{equation}
So at any time $t \in (0,1)$ the positions of $n$ non-intersecting
Brownian bridges starting at $0$ and ending at specified points are
distributed as the eigenvalues of a Gaussian random matrix with
external source.

The connection between random matrices and non-intersecting random paths
is actually well-known, see e.g.\ the recent works \cite{Baik}, \cite{Joh},
\cite{KT}, \cite{NF} and references cited therein.
\medskip

P.~Zinn-Justin showed that the $m$-point correlation functions for the
eigenvalues of $M$ have determinantal form
\begin{equation}\label{corfunctions}
R_m(\la_1,\ldots,\la_m)=\det\left(K_n(\la_j,\la_k)\right)_{1\le j,k\le m}.
\end{equation}
It was shown in \cite{BK1} that the average characteristic
polynomial
\[ P(z) = \mathbb E \left[ \det (zI - M) \right] \]
is a multiple orthogonal polynomial of type II, which for the Gaussian
case (\ref{gausspotential}) is a multiple Hermite polynomial, see \cite{ABVA},
\cite{BK3}, and that the correlation kernel $K_n$ can be expressed in
terms of the solution of the RH problem for multiple orthogonal
polynomials \cite{VAGK}, see also \cite{DK}.

We state the RH problem here for the Gaussian case (\ref{gausspotential})
and for the external source (\ref{extsource2}) where $n$ is even.
Then the RH problem asks for a $3 \times 3$ matrix valued function $Y$
satisfying the following.
\begin{itemize}
\item $Y : \mathbb C \setminus \mathbb R \to \mathbb C^{3\times 3}$ is analytic.
\item For $x \in \mathbb R$, there is a jump
\begin{equation} \label{RH-Y2}
    Y_+(x) = Y_-(x) \begin{pmatrix}
    1 & w_1(x) & w_2(x) \\
    0 & 1 & 0 \\
    0 & 0 & 1 \end{pmatrix}
\end{equation}
where
\begin{equation} \label{weights}
    w_1(x) = e^{-n(x^2/2-ax)}, \qquad
    w_2(x) = e^{-n(x^2/2+ax)},
\end{equation}
and $Y_+(x)$ ($Y_-(x)$) denotes the limit of $Y(z)$ as $z \to x \in \mathbb R$
from the upper (lower) half-plane.
\item As $z \to \infty$, we have
\begin{equation} \label{RH-Y3}
    Y(z) = (I + O(1/z)) \begin{pmatrix}
    z^{n} & 0 & 0 \\
    0 & z^{-n/2} & 0 \\
    0 & 0 & z^{-n/2} \end{pmatrix}.
    \end{equation}
\end{itemize}
The RH problem has a unique solution in terms of
multiple Hermite polynomials and their Cauchy transforms \cite{BK1}, \cite{VAGK}.
The correlation kernel $K_n$ is expressed in terms of $Y$ as follows
\begin{equation} \label{kernelK}
    K_n(x,y) = \frac{e^{-\frac{1}{4}n(x^2+y^2)}}{2\pi i (x-y)}
    \begin{pmatrix} 0 & e^{nay} & e^{-nay} \end{pmatrix}
    Y^{-1}(y) Y(x)
    \begin{pmatrix} 1 \\ 0 \\ 0 \end{pmatrix}
\end{equation}.

Our goal is to analyze the above RH problem in the large $n$ limit
and to obtain from this scaling limits of the
kernel (\ref{kernelK}) in various regimes.
In this paper we consider the case $0 < a < 1$.
The case $a > 1$ was considered in \cite{BK2} and the critical case $a=1$
will be considered in \cite{BK4}. First we describe the limiting mean
density of eigenvalues.

\begin{theo} \label{maintheo1}
The limiting mean density of eigenvalues
\begin{equation} \label{rho1}
 \rho(x) = \lim_{n \to \infty} \frac{1}{n} K_n(x,x)
\end{equation}
exists for every $a > 0$. It satisfies
\begin{equation} \label{pastur1}
    \rho(x)=\frac{1}{\pi} \left|\, \Im\xi(x)\right|,
\end{equation}
where $\xi=\xi(x)$ is a solution of the cubic equation,
\begin{equation} \label{pastur2}
    \xi^3-x\xi^2-(a^2-1)\xi+xa^2=0.
\end{equation}
The support of $\rho$ consists of those $x \in \mathbb R$
for which {\rm (\ref{pastur2})} has a non-real solution.
\begin{enumerate}
\item[\rm (a)] For $0 < a < 1$,
the support of $\rho$ consists of one interval
$[-z_1, z_1]$,
and $\rho$ is real analytic and positive on $(-z_1, z_1)$, and it
vanishes like a square root at the edge points $\pm z_1$, i.e.,
there exists a constant $\rho_1 > 0$ such that
\begin{equation} \label{density1}
    \rho(x) = \frac{\rho_1}{\pi} |x \mp z_1|^{1/2} (1+o(1))
    \qquad \mbox{as } x \to \pm z_1, \, x \in (-z_1, z_1).
\end{equation}
\item[\rm (b)] For $a=1$,
the support of $\rho$ consists of one interval
$[-z_1, z_1]$,
and $\rho$ is real analytic and positive on $(-z_1, 0) \cup (0, z_1)$, it
vanishes like a square root at the edge points $\pm z_1$, and it
vanishes like a third root at $0$, i.e., there exists a constant $c > 0$
such that
\begin{equation} \label{density2}
    \rho(x) = c |x|^{1/3} \left(1+ o(1) \right), \qquad
        \mbox{as } x \to 0.
\end{equation}
\item[\rm (c)] For $a > 1$,
the support of $\rho$ consists of two disjoint intervals
$[-z_1, -z_2] \cup [z_2, z_1]$ with $0 < z_2 < z_1$,
$\rho$ is real analytic and positive on $(-z_1, -z_2) \cup (z_2, z_1)$,
and it vanishes like a square root at the edge points $\pm z_1$, $\pm z_2$.
\end{enumerate}
\end{theo}

\begin{remark}
Theorem \ref{maintheo1} is a very special case of a theorem
of Pastur \cite{Pas} on the eigenvalues of a matrix $M = M_0 + A$
where $M_0$ is random and $A$ is deterministic as in (\ref{randomplusdet}).
Since in this paper our interest is in the case $0 < a < 1$, we
show in Section 9 how Theorem \ref{maintheo1} follows from
our methods for this case. See \cite{BK2} for the case $a > 1$.
\end{remark}

\begin{remark}
Theorem \ref{maintheo1} has the following interpretation
in terms of non-intersecting Brownian motions starting at $0$
and ending at some specified points $b_j$. We suppose $n$ is even,
and we let half of the $b_j$'s coincide with $b > 0$ and the other
half with $-b$. Then as explained before, at time $t \in (0,1)$
the (rescaled) positions of the Brownian paths coincide with the
eigenvalues of the Gaussian random matrix with external
source (\ref{extsource2}) where
\begin{equation} \label{effa}
    a = b \sqrt\frac{t}{1-t}.
\end{equation}
The phase transition at $a=1$ corresponds to
\[ t = t_c \equiv \frac{1}{1+b^2}. \]
So, by Theorem \ref{maintheo1}, the limiting distribution of
the Brownian paths as $n \to\infty$ is supported by one interval
when $t < t_c$ and by two intervals when $t > t_c$. At the
critical time $t_c$ the two groups of Brownian paths split,
with one group ending at $t =1$ at $b$ and the other at $-b$.
\end{remark}

As in \cite{BK2} we formulate our main result in terms
of a rescaled version of the kernel $K_n$
\begin{equation} \label{rescaledkernel}
\hat{K}_n(x,y) = e^{n(h(x)-h(y))} K_n(x,y)
\end{equation}
for some function $h$. The rescaling (\ref{rescaledkernel})
does not affect the correlation functions (\ref{corfunctions}).

\begin{theo} \label{maintheo2}
Let $0 < a < 1$ and let $z_1$ and $\rho$ be as in Theorem {\rm \ref{maintheo1} (a)}.
Then there is a function $h$ such that the following hold for
the rescaled kernel {\rm (\ref{rescaledkernel})}.
\begin{enumerate}
\item[\rm (a)] For every
$x_0 \in (-z_1, z_1)$ and $u,v \in \R$, we have
\begin{equation} \label{universalbulk}
    \lim_{n\to\infty}
    \frac{1}{n\rho(x_0)}
    \hat{K}_n \left(x_0 + \frac{u}{n\rho(x_0)}, x_0 + \frac{v}{n \rho(x_0)}\right)
    = \frac{\sin  \pi(u-v)}{\pi(u-v)}.
    \end{equation}
\item[\rm (b)] For every $u, v \in \mathbb R$ we have
\begin{equation} \label{universaledge1}
\lim_{n \to \infty}
    \frac{1}{(\rho_1 n)^{2/3}}
    \hat{K}_n\left(z_1 + \frac{u}{(\rho_1 n)^{2/3}}, z_1 + \frac{v}{(\rho_1 n)^{2/3}}\right)
    = \frac{\Ai(u)\Ai'(v)-\Ai'(u)\Ai(v)}{u-v},
    \end{equation}
where $\Ai$ is the usual Airy function, and  $\rho_1$ is the constant from
{\rm (\ref{density1})}.
\end{enumerate}
\end{theo}

Theorem \ref{maintheo2} is similar to the main theorems Theorem 1.2
and Theorem 1.3 of \cite{BK2}. It expresses that the local eigenvalue correlations
show the universal behavior as $n \to \infty$, both in the bulk and at the edge,
that is well-known from unitary random matrix models. So the result itself is
not that surprising.

To obtain Theorem \ref{maintheo2} we use the Deift/Zhou steepest descent
method for RH problems and a main tool
is the three-sheeted Riemann surface associated with equation (\ref{pastur2})
as in \cite{BK2}.  There is however an
important technical difference with \cite{BK2}. For $a > 1$, the branch
points of the Riemann surface are all real, and they correspond to
the four edge points $\pm z_1$, $\pm z_2$ of the support as described in Theorem
\ref{maintheo1} (c). For $a < 1$, two branch points are purely imaginary
and they have no direct meaning for the problem at hand. The other two branch points
are real and they correspond to the edge points $\pm z_1$ as in Theorem
\ref{maintheo1} (a). See Figure 1 for the sheet structure of the
Riemann surface. The branch points on the non-physical sheets result in
a non-trivial modification of the steepest descent method.
As already mentioned before, one of the steps involves a global opening of lenses,
and this is the main new technical contribution of this paper.

\medskip

The rest of the paper is devoted to the proof of the theorems
with the Deift/Zhou steepest descent method for RH problems.
It consists of a sequence of transformations which reduce
the original RH problem to
a RH problem which is normalized at infinity, and whose
jump matrices are uniformly close to the identity as $n \to \infty$.
In this paper there are four transformations  $Y \mapsto U \mapsto T
\mapsto S \mapsto R$. A main role is played by the Riemann surface (\ref{pastur2})
and certain $\lambda$-functions defined on it. These
are introduced in the next section, and they are used
in Section 3 to define the first transformation $Y \mapsto U$.
This transformation has the effect of normalizing the RH problem
at infinity, and in addition of producing ``good'' jump matrices
that are amenable to subsequent analysis. However, contrary
to earlier works, some of the jump matrices for $U$ have
entries that are exponentially growing as $n \to \infty$.
These exponentially growing entries disappear after the
second transformation $U \mapsto T$ in Section 4 which
involves the global opening of lenses.

The remaining transformation follow the pattern of \cite{BK2},
\cite{DKMVZ1}, \cite{DKMVZ2} and other works.
The transformation $T \mapsto S$ in Section 5 involves a local opening of
lenses which turns the remaining oscillating entries into
exponentially decaying ones. Then a parametrix for $S$ is built
in Sections 6 and 7. In Section 6 a model RH problem is solved
which provides the parametrix for $S$ away from the branch points,
and in Section 7 local parametrices are built around each of the branch
points with the aid of Airy functions.
Using this parametrix we define the final transformation $S \mapsto R$
in Section 8. It leads to a RH problem for $R$ which is of the desired
type:  normalized at infinity and jump matrices tending to the identity
as $n \to \infty$. Then $R$ itself tends to the identity matrix
as $n \to \infty$, which is then used in the final Section 9
to prove the Theorems \ref{maintheo1} and \ref{maintheo2}.

\section{The Riemann surface and $\lambda$-functions}
We start from the cubic equation (\ref{pastur2}) which we write
now with the variable $z$ instead of $x$
\begin{equation} \label{cubicequation}
    \xi^3 - z \xi^2 + (1-a^2) \xi + z a^2 = 0.
\end{equation}
It defines a Riemann surface that will play a central
role in the proof. The inverse mapping is given by  the rational function
\begin{equation} \label{mapping}
    z = \frac{\xi^3-(a^2-1) \xi}{\xi^2-a^2}.
\end{equation}
There are four branch points $\pm z_1$, $\pm i z_2$
with $z_1 > z_2 > 0$, which can be found as the images of
the critical points under the inverse mapping.
The mapping (\ref{mapping}) has three inverses,  $\xi_j(z)$, $j=1,2,3$,
that behave near infinity as
\begin{equation} \label{psiatinfinity}
\begin{aligned}
    \xi_1(z) & =  z - \frac{1}{z} + O(1/z^2), \\
    \xi_2(z) & =  a + \frac{1}{2z} + O(1/z^2), \\
    \xi_3(z) & =  -a + \frac{1}{2z} + O(1/z^2).
\end{aligned}
\end{equation}
The sheet structure of the Riemann surface is determined by
the way we choose the analytical continuations of the $\xi_j$'s.

It  may be checked that $\xi_1$ has an analytic continuation to
$\mathbb C \setminus [-z_1,z_1]$, which we take as the first sheet.
The functions $\xi_2$ and $\xi_3$ have analytic continuations to
$\mathbb C \setminus \setminus ([0,z_1] \cup [-iz_2, iz_2])$
and $\mathbb C \setminus ([-z_1,0] \cup [-iz_2, iz_2])$, respectively,
which we take to be the second and third sheets, respectively.
So the second and third sheet are connected along $[-iz_2,iz_2]$, the
first sheet is connected with the second sheet along $[0, z_1]$,
and the first sheet is connected with the third sheet along $[-z_1,0]$,
see Figure 1.
\begin{center}
 \begin{figure}[h]\label{figure1}
\begin{center}
   \scalebox{0.5}{\includegraphics{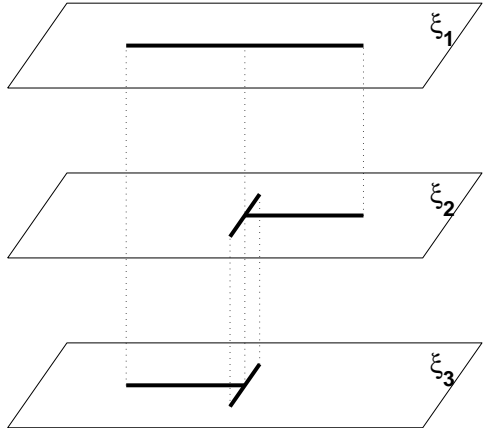}}
\end{center}
        \caption{The Riemann surface $\xi^3 - z \xi^2 + (1-a^2) \xi + z a^2 = 0$.}
    \end{figure}
\end{center}
We note the jump relations
\begin{equation} \label{jumpsxi}
\begin{array}{ll}
    \xi_{1\mp} = \xi_{2\pm} & \textrm{ on } (0,z_1), \\
    \xi_{1\mp} = \xi_{3\pm} & \textrm{ on } (-z_1,0),  \\
    \xi_{2\mp} = \xi_{3\pm} & \textrm{ on } (-iz_2,iz_2).
\end{array}
\end{equation}

The $\lambda$-functions are primitives of the $\xi$-functions
$\lambda_j(z) = \int^z \xi_j(s) ds$, more precisely
\begin{equation} \label{deflambda}
\begin{aligned}
    \lambda_1(z) &= \int_{z_1}^z \xi_1(s) ds \\
    \lambda_2(z) &= \int_{z_1}^z \xi_2(s) ds \\
    \lambda_3(z) &= \int_{-z_{1+}}^z \xi_3(s) ds + \lambda_{1-}(-z_1)
\end{aligned}
\end{equation}
The path of integration for $\lambda_3$ lies in $\mathbb C \setminus
((-\infty,0] \cup [-iz_2,iz_2])$, and it starts at the point $-z_1$
on the upper side of the cut.
All three $\lambda$-functions are defined on their respective sheets of
the Riemann surface with an additional cut along the negative real
axis. Thus $\lambda_1, \lambda_2, \lambda_3$ are defined and analytic
on $\mathbb C \setminus (-\infty, z_1]$, $\mathbb C \setminus
((-\infty,z_1] \cup [-iz_2,iz_2])$, and $\mathbb C \setminus
((-\infty,0] \cup [-iz_2,iz_2])$, respectively.
Their behavior at infinity is
\begin{equation} \label{lambdaatinfinity}
\begin{aligned}
    \lambda_1(z) &= \frac{1}{2}z^2 - \log z + \ell_1 + O(1/z) \\
    \lambda_2(z) &= az + \frac{1}{2} \log z + \ell_2 + O(1/z) \\
    \lambda_3(z) &= -az + \frac{1}{2} \log z + \ell_3 + O(1/z)
\end{aligned}
\end{equation}
for certain constants $\ell_j$, $j=1,2,3$.
The $\lambda_j$'s satisfy the following jump relations
\begin{equation} \label{jumpslambda}
\begin{array}{ll}
    \lambda_{1 \mp}  = \lambda_{2\pm} &\textrm{ on } (0,z_1), \\
    \lambda_{1-}  = \lambda_{3+}  & \textrm{ on } (-z_1,0), \\
    \lambda_{1+}  = \lambda_{3-} -\pi i & \textrm{ on } (-z_1,0), \\
    \lambda_{2\mp}  = \lambda_{3\pm}  & \textrm{ on } (0, iz_2), \\
    \lambda_{2\mp}  = \lambda_{3\pm} - \pi i & \textrm{ on } (-iz_2,0), \\
    \lambda_{1+} = \lambda_{1-} - 2\pi i & \textrm{ on } (-\infty, -z_1), \\
    \lambda_{2+} = \lambda_{2-} + \pi i & \textrm{ on } (-\infty, 0), \\
    \lambda_{3+} = \lambda_{3-} + \pi i & \textrm{ on } (-\infty, -z_1),
\end{array}
\end{equation}
where the segment $(-iz_2, iz_2)$ is oriented upwards. We obtain
(\ref{jumpslambda}) from (\ref{jumpsxi}), (\ref{deflambda}), and
the values of the contour integrals around the cuts in the positive
direction
\[
    \oint \xi_1(s) ds  = - 2\pi i, \qquad
    \oint \xi_2(s) ds  = \pi i, \qquad
    \oint \xi_3(s) ds  = \pi i,
\]
which follow from (\ref{psiatinfinity}).

\begin{remark}
We have chosen the segment $[-iz_2, iz_2]$ as the cut that connects
the branch points $\pm iz_2$. We made this choice because of symmetry
and ease of notation, but it is not essential. Instead
we could have taken an arbitrary smooth curve lying in
the region bounded by the four smooth curves in Figure 2 (see the next section)
that connect the points $x_0, iz_2, -x_0$, and $-iz_2$. For any such
curve, the subsequent analysis would go through without any
additional difficulty.
\end{remark}

\section{First transformation $Y \mapsto U$}
We define for $z \in \mathbb C \setminus (\mathbb R \cup [-iz_2, iz_2])$,
\begin{equation} \label{defU}
    U(z) = \diag(e^{-n\ell_1}, e^{-n \ell_2}, e^{-n\ell_3})
    Y(z) \diag(e^{n (\lambda_1(z) - \frac{1}{2} z^2)}, e^{n(\lambda_2(z) - az)},
        e^{n(\lambda_3(z) + az)}).
\end{equation}
This coincides with the first transformation in \cite{BK2}.
Then $U$ solves the following RH problem.

\begin{itemize}
\item $U : \mathbb C \setminus (\mathbb R \cup [-iz_2, iz_2]) \to \mathbb C^{3 \times 3}$ is analytic.
\item $U$ satisfies the jumps
\begin{equation} \label{jumpU0}
    U_+ = U_-
    \begin{pmatrix}
        e^{n(\lambda_{1+}-\lambda_{1-})} & e^{n(\lambda_{2+}-\lambda_{1-})}
        & e^{n(\lambda_{3+}-\lambda_{1-})} \\
        0 & e^{n(\lambda_{2+}-\lambda_{2-})} & 0 \\
        0 & 0 & e^{n(\lambda_{3+}-\lambda_{3-})}
        \end{pmatrix}
        \qquad \textrm{ on } \mathbb R,
\end{equation}
and
\begin{equation}
   \label{jumpU5}
    U_+ = U_-
    \begin{pmatrix}
        1 & 0 & 0 \\
        0 & e^{n(\lambda_{2+}-\lambda_{2-})} & 0 \\
        0 & 0 & e^{n(\lambda_{3+}-\lambda_{3-})}
    \end{pmatrix}
    \qquad \textrm{on } [-iz_2,iz_2].
\end{equation}
\item $U(z) = I + O(1/z)$ as $z \to \infty$.
\end{itemize}
The asymptotic condition follows from (\ref{RH-Y3}), (\ref{lambdaatinfinity})
and the definition of $U$.

The jump on the real line (\ref{jumpU0}) takes on a different form on the four
intervals $(-\infty, -z_1]$, $[-z_1, 0)$, $(0, z_1]$,
and $[z_1, \infty)$. Indeed we get from (\ref{jumpslambda}), (\ref{jumpU0}),
and the fact that $n$ is even,
\begin{eqnarray} \label{jumpU1}
    U_+ & = & U_-
    \begin{pmatrix}
        1 & e^{n(\lambda_{2+}- \lambda_{1-})} & e^{n(\lambda_{3+}-\lambda_{1-})} \\
        0 & 1 & 0 \\
        0 & 0 & 1
        \end{pmatrix}  \qquad \textrm{on } (-\infty,-z_1] \\
    \label{jumpU2}
        U_+ & = & U_-
    \begin{pmatrix}
        e^{n(\lambda_{1+}-\lambda_{1-})} & e^{n(\lambda_{2+}-\lambda_{1-})} & 1 \\
        0 & 1 & 0 \\
        0 & 0 & e^{n(\lambda_{3+}-\lambda_{3-})}
        \end{pmatrix}
        \qquad \textrm{on } (-z_1,0) \\
    \label{jumpU3}
    U_+ & = & U_-
    \begin{pmatrix}
        e^{n(\lambda_{1+}-\lambda_{1-})} & 1 & e^{n(\lambda_3-\lambda_{1-})} \\
        0 & e^{n(\lambda_{2+}-\lambda_{2-})} & 0 \\
        0 & 0 & 1
    \end{pmatrix}
    \qquad \textrm{on } (0, z_1)  \\
  \label{jumpU4} U_+ & = & U_-
    \begin{pmatrix}
        1 & e^{n(\lambda_2-\lambda_1)} & e^{n(\lambda_3-\lambda_1)} \\
        0 & 1 & 0 \\
        0 & 0 & 1
        \end{pmatrix}
        \qquad \textrm{on } [z_1, \infty).
\end{eqnarray}

Now to see what has happened it is important to know
the sign of $\Re (\lambda_j - \lambda_k)$ for $j \neq k$.
Figure~2 shows the curves where $\Re \lambda_j = \Re \lambda_k$.

  \begin{figure}[h] \label{figure2}
  \centerline{
  \includegraphics[width=12cm]{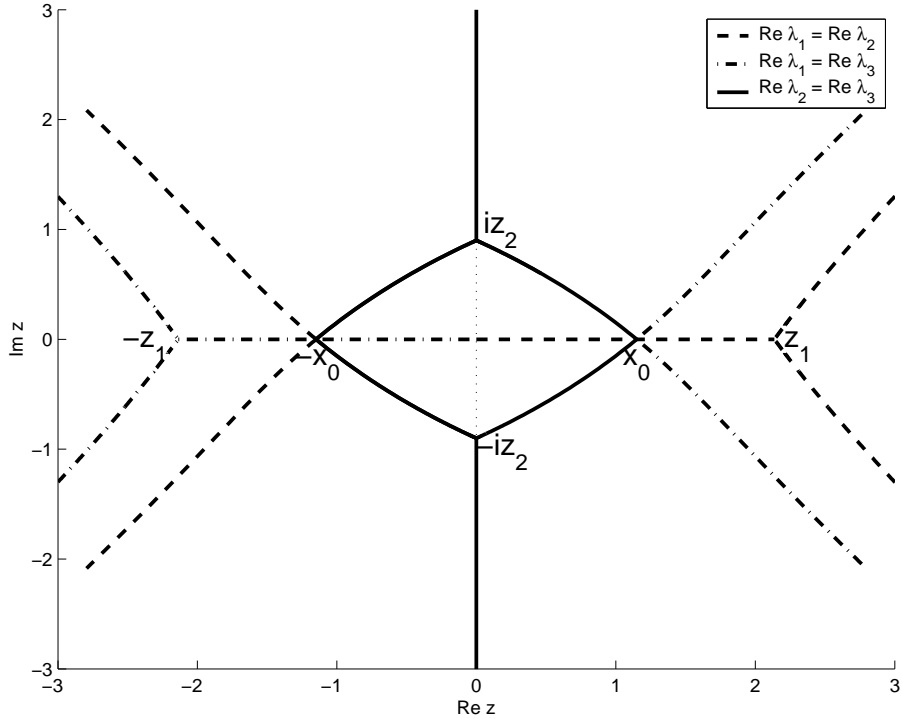}}
        \caption{Curves where $\Re \lambda_1 = \Re \lambda_2$ (dashed lines),
        $\Re \lambda_1 = \Re \lambda_3$ (dashed-dotted lines), and
        $\Re \lambda_2 = \Re \lambda_3$ (solid lines). This particular
        figure is for the value $a=0.4$.
        }
    \end{figure}

From each of the branch points $\pm z_1$, $\pm i z_2$ there are three
curves emanating at equal angle of $2 \pi/3$.
We have $\Re \lambda_1 = \Re \lambda_2$ on the interval
$[0, z_1]$ and on two unbounded curves from $z_1$.
Similarly, $\Re \lambda_1 = \Re \lambda_3$ on the interval
$[-z_1,0]$ and on two unbounded curves from $-z_1$.
We have $\Re \lambda_2 = \Re \lambda_3$ on the curves that
emanate from  $\pm i z_2$. That is, on the vertical half-lines
$[iz_2, +i\infty)$ and $(-i\infty, -iz_2]$ and on four other
curves, before they intersect the real axis.
The points where they intersect the real axis are $\pm x_0$
for some $x_0 \in (0, z_1)$. After that point we have
$\Re \lambda_1 = \Re \lambda_3$ for the curves in the
right half-plane and $\Re \lambda_1 = \Re \lambda_2$
for the curves in the left half-plane.
Figure 2 was produced with Matlab for the value $a=0.4$.
The picture is similar for other values of $a \in (0,1)$.
As $a \to 0+$ or $a \to 1-$, the imaginary branch points $\pm iz_2$
tend to the origin.

Using Figure 2 and the asymptotic behavior (\ref{lambdaatinfinity}) we
can determine the ordering of $\Re \lambda_j$, $j=1,2,3$
in every domain in the plane. Indeed, in the domain on the right,
bounded by the two unbounded curves emanating from $z_1$, we have
$\Re \lambda_1 > \Re \lambda_2 > \Re \lambda_3$ because of (\ref{lambdaatinfinity}).
Then if we go to a neighboring domain, we pass a curve where
$\Re \lambda_1 = \Re \lambda_2$, and so the ordering changes to
$\Re \lambda_2 > \Re \lambda_1 > \Re \lambda_3$. Continuing in
this way, and also taking into account the cuts that we have for
the $\lambda_j$'s,  we find the ordering in any domain.

Inspecting the jump matrices for $U$ in (\ref{jumpU5})--(\ref{jumpU4}), we
then find the following:
\begin{enumerate}
\item[(a)] The non-zero off-diagonal entries in the jump matrices in
(\ref{jumpU1}) and (\ref{jumpU4}) are exponentially small, and the
jump matrices tend to the identity matrix as $n \to \infty$.
\item[(b)] The non-constant diagonal entries in the jump matrices in (\ref{jumpU2})
and (\ref{jumpU3}) have modulus one, and they are rapidly oscillating
for large $n$.
\item[(c)] The $(1,2)$-entry in the jump matrix in (\ref{jumpU2})
is exponentially decreasing on $(-z_1,-x_0)$, but exponentially
increasing on $(-x_0,0)$ as $n \to \infty$. Similarly, the $(1,3)$-entry in
the jump matrix in (\ref{jumpU3})
is exponentially decreasing on $(x_0, z_1)$, and exponentially
increasing on $(0,x_0)$.
\item[(d)] The entries in the jump matrix in (\ref{jumpU5})
are real. The $(2,2)$-entry is exponentially increasing as $n \to \infty$,
and the $(3,3)$-entry is exponentially decreasing.
\end{enumerate}

The exponentially increasing entries observed in items (c) and (d) are
undesirable, and this might lead to the impression that the
first transformation $Y \mapsto U$ was not the right thing to do.
However, after the second transformation which we do in the next section,
all exponentially increasing entries miraculously disappear.

\section{Second transformation $U \mapsto T$}

  \begin{figure}[h] \label{figure3}
  \centerline{
  \includegraphics[width=12cm]{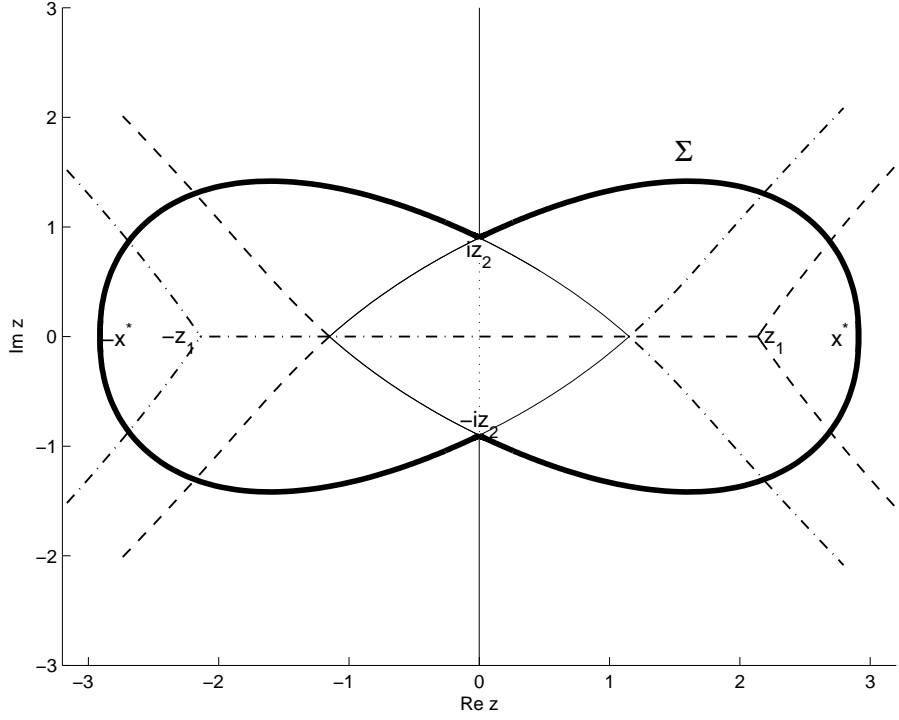}}
        \caption{Contour $\Sigma$ which is such that $\Re \lambda_2 < \Re \lambda_3$
        on the part of $\Sigma$ in the left half-plane and $\Re \lambda_2 > \Re \lambda_3$
        on the part of $\Sigma$ in the right half-plane.
        }
    \end{figure}

The second transformation involves the global opening of lenses already
mentioned in the introduction. It is needed to turn the exponentially
increasing entries in the jump matrices into exponentially decreasing ones.

Let $\Sigma$ be a closed curve, consisting of a part in
the left half-plane from $-iz_2$ to $iz_2$, symmetric with respect
to the real axis, plus its mirror image in the right
half-plane.  The part in the left half-plane lies entirely in the region
where $\Re \lambda_2 < \Re \lambda_3$ and it intersects
the negative real axis in a point $-x^*$ with $x^* > z_1$, see Figure 3.
So $\Sigma$ avoids the region bounded by the curves from $\pm i z_2$ to
$\pm x_0$.  In a neighborhood of $iz_2$ we take $\Sigma$
to be the analytic continuation
of the curves where $\Re \lambda_2 = \Re \lambda_3$. As a result, this means
that
\begin{equation} \label{constrSigma}
    \lambda_2 - \lambda_3  \quad \textrm{is real on } \Sigma
    \textrm{ in a neighborhood of } iz_2.
\end{equation}
This will be convenient
for the construction of the local parametrix in Section 7.

The contour $\Sigma$ encloses a bounded domain and we make the
second transformation in that domain only.
So we put $T = U$ outside $\Sigma$ and inside $\Sigma$ we put
\begin{equation} \label{defT}
\begin{aligned}
    T& = U \begin{pmatrix}
    1 & 0 & 0 \\
    0 & 1 & 0 \\
    0 & - e^{n(\lambda_2-\lambda_3)} & 1 \end{pmatrix}
    \qquad \textrm{for } \Re z < 0 \textrm{ inside } \Sigma, \\
    T&= U \begin{pmatrix}
    1 & 0 & 0 \\
    0 & 1 & - e^{n(\lambda_3-\lambda_2)} \\
    0 & 0 & 1 \end{pmatrix}
    \qquad \textrm{for } \Re z > 0 \textrm{ inside } \Sigma.
\end{aligned}
\end{equation}

  \begin{figure}[h] \label{figure4}
  \centerline{
  \includegraphics[width=16cm,height=8cm]{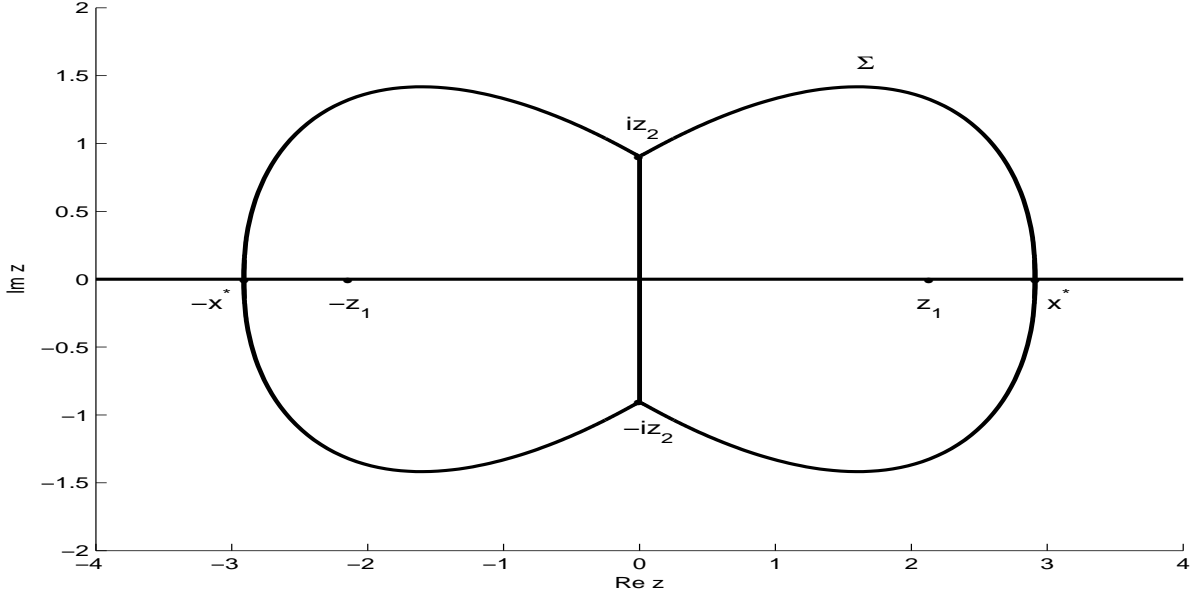}}
        \caption{$T$ has jumps on the  real line, the interval $[-iz_2,iz_2]$
        and on $\Sigma$.}
    \end{figure}

Then $T$ is defined and analytic outside the contours shown in Figure~4.
Using the jumps for $U$ and the definition (\ref{defT}) we
calculate the jumps for $T$ on any part of the contour. We get different
expressions for six real intervals, for the vertical segment $[-iz_2,iz_2]$,
and for $\Sigma$ (oriented clockwise) in the left and right half-planes.
The result is that $T$ satisfies the following RH problem.

\begin{itemize}
\item $T : \mathbb C \setminus (\mathbb R \cup [-iz_2,iz_2] \cup \Sigma) \to \mathbb C^{3 \times 3}$ is analytic.
\item $T$ satisfies the following jump relations on the real line
\begin{align} \label{jumpT1}
    T_+ & = T_-
    \begin{pmatrix}
        1 & e^{n(\lambda_{2+}- \lambda_{1-})} & e^{n(\lambda_{3+}-\lambda_{1-})} \\
        0 & 1 & 0 \\
        0 & 0 & 1
        \end{pmatrix}  \qquad \textrm{on } (-\infty,-x^*] \\
    \label{jumpT2}
        T_+ & = T_-
    \begin{pmatrix}
        1 & 0 & e^{n(\lambda_{3+}-\lambda_{1-})} \\
        0 & 1 & 0 \\
        0 & 0 & 1
        \end{pmatrix} \qquad \textrm{on } (-x^*, -z_1] \\
    \label{jumpT3}
        T_+ & = T_-
    \begin{pmatrix}
        e^{n(\lambda_{1+}-\lambda_{1-})} & 0 & 1 \\
        0 & 1 & 0 \\
        0 & 0 & e^{n(\lambda_{3+}-\lambda_{3-})}
        \end{pmatrix}
        \qquad \textrm{on } (-z_1,0) \\
    \label{jumpT4}
    T_+ & = T_-
    \begin{pmatrix}
        e^{n(\lambda_{1+}-\lambda_{1-})} & 1 & 0 \\
        0 & e^{n(\lambda_{2+}-\lambda_{2-})} & 0 \\
        0 & 0 & 1
    \end{pmatrix}
    \qquad \textrm{on } (0, z_1)  \\
      \label{jumpT5}
        T_+ & = T_-
      \begin{pmatrix}
        1 & e^{n(\lambda_2-\lambda_1)} & 0 \\
        0 & 1 & 0 \\
        0 & 0 & 1
        \end{pmatrix}
        \qquad \textrm{on } [z_1, x^*) \\
    \label{jumpT6}
        T_+ & = T_-
      \begin{pmatrix}
        1 & e^{n(\lambda_2-\lambda_1)} & e^{n(\lambda_3-\lambda_1)} \\
        0 & 1 & 0 \\
        0 & 0 & 1
        \end{pmatrix}
        \qquad \textrm{on } [z_1, \infty).
\end{align}
The jump on the vertical segment is
\begin{align}
   \label{jumpT7}
    T_+ & = T_-
    \begin{pmatrix}
        1 & 0 & 0 \\
        0 & 0 & 1 \\
        0 & -1 & e^{n(\lambda_{3+}-\lambda_{3-})}
    \end{pmatrix}
    \qquad \textrm{on } [-iz_2,iz_2].
\end{align}
The jumps on $\Sigma$ are
\begin{align}
    \label{jumpT8}
    T_+ & = T_-
    \begin{pmatrix}
    1 & 0 & 0 \\
    0 & 1 & 0 \\
    0 & e^{n(\lambda_2-\lambda_3)} & 1
    \end{pmatrix}
    \qquad \textrm{on } \{ z \in \Sigma \mid \Re z < 0 \} \\
    \label{jumpT9}
    T_+ & = T_-
    \begin{pmatrix}
    1 & 0 & 0 \\
    0 & 1 & e^{n(\lambda_3-\lambda_2)} \\
    0 & 0 & 1
    \end{pmatrix}
    \qquad \textrm{on } \{ z \in \Sigma \mid \Re z > 0 \}.
\end{align}
\item $T(z) = I + O(1/z)$ as $z \to \infty$.
\end{itemize}

Now the jump matrices are nice. Because of our choice of $\Sigma$
we have that the jump matrices in (\ref{jumpT8}) and (\ref{jumpT9})
converge to the identity matrix as $n \to \infty$.
Also the jump matrices in (\ref{jumpT1}), (\ref{jumpT2}), (\ref{jumpT5})
and (\ref{jumpT6}) converge to the identity matrix as $n \to \infty$.
The $(3,3)$-entry in the jump matrix in (\ref{jumpT7}) is exponentially
small, so that this matrix tends to $\begin{pmatrix} 1 & 0 & 0 \\
0 & 0 & 1 \\ 0 & -1 & 0 \end{pmatrix}$.

The jump matrices in (\ref{jumpT4}) and (\ref{jumpT5}) have
oscillatory entries on the diagonal, and they are turned into
exponential decaying off-diagonal entries by opening a (local) lens
around $(-z_1, z_1)$. This is the next transformation.

\section{Third transformation $T \mapsto S$}
We are now going to open up a lens around $(z_1, z_1)$ as in Figure 5.
There is no need to treat $0$ as a special point.

  \begin{figure}[h] \label{figure5}
  \centerline{
  \includegraphics[width=16cm,height=8cm]{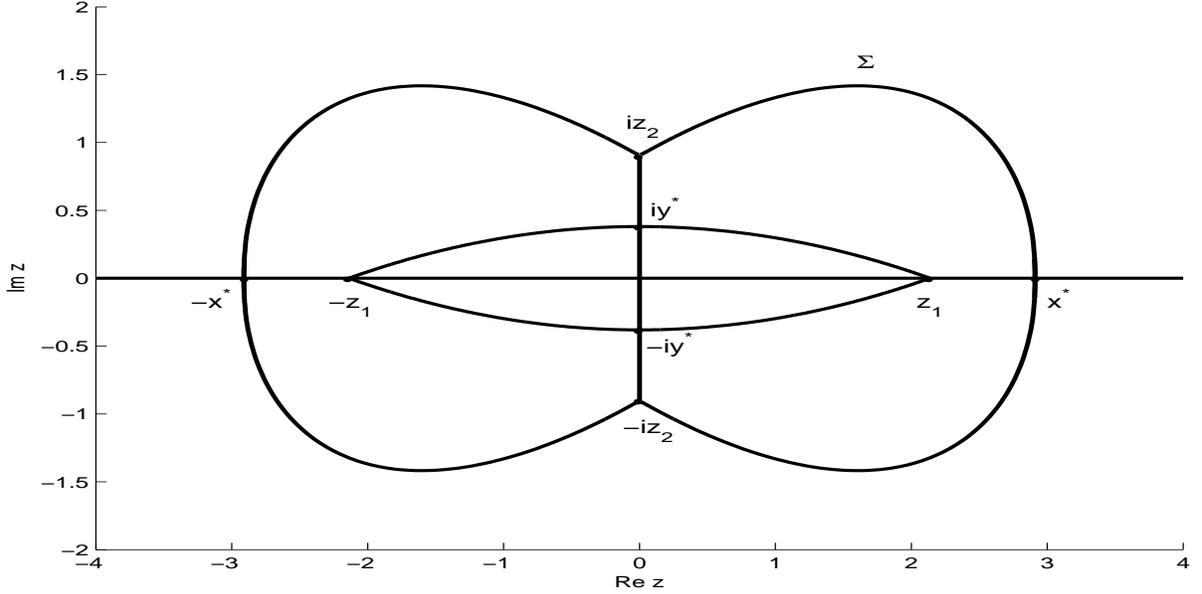}}
        \caption{Opening of lens around $[-z_1,z_1]$. The
        new matrix-valued function $S$ has jumps on the  real line,
        the interval $[-iz_2,iz_2]$, on $\Sigma$,
        and on the upper and lower lips of the lens around $[-z_1,z_1]$.}
    \end{figure}

The jump matrix on $(-z_1,0)$, see (\ref{jumpT3}), has factorization
\begin{equation} \label{factorization1}
\begin{aligned} T_-^{-1}T_+ & =
    \begin{pmatrix}
        e^{n(\lambda_1-\lambda_3)_+} & 0 & 1 \\
        0 & 1 & 0 \\
        0 & 0 & e^{n(\lambda_1-\lambda_3)_-}
        \end{pmatrix} \\
      & =
      \begin{pmatrix}
      1 & 0 & 0 \\ 0 & 1 & 0 \\
      e^{n(\lambda_1-\lambda_3)_-} & 0 & 1 \end{pmatrix}
      \begin{pmatrix}
      0 & 0 & 1 \\ 0 & 1 & 0 \\ -1 & 0 & 0 \end{pmatrix}
      \begin{pmatrix}
      1 & 0 & 0 \\ 0 & 1 & 0 \\
      e^{n(\lambda_1-\lambda_3)_+} & 0 & 1 \end{pmatrix}
\end{aligned}
\end{equation}
and the jump matrix on $(0,z_1)$, see (\ref{jumpT4}), has
factorization
\begin{equation} \label{factorization2}
\begin{aligned}
   T_-^{-1} T_+ & =
    \begin{pmatrix}
        e^{n(\lambda_1-\lambda_2)_+} & 1 & 0 \\
        0 & e^{n(\lambda_1-\lambda_2)_-} & 0 \\
        0 & 0 & 1
    \end{pmatrix} \\
    & =
      \begin{pmatrix}
      1 & 0 & 0 \\ e^{n(\lambda_1-\lambda_2)_-} & 1 & 0 \\
      0 & 0 & 1 \end{pmatrix}
      \begin{pmatrix}
      0 & 1 & 0 \\ -1 & 0 & 0 \\ 0 & 0 & 1 \end{pmatrix}
      \begin{pmatrix}
      1 & 0 & 0 \\ e^{n(\lambda_1-\lambda_2)_+} & 1 & 0 \\
      0 & 0 & 1 \end{pmatrix}
\end{aligned}
\end{equation}

We open up the lens on $[-z_1,z_1]$ and we make sure that it stays inside
$\Sigma$. We assume that the lens is symmetric with respect to the real
and imaginary axis. The point where the upper lip intersects the imaginary axis
is called $i y^*$. Then we define $S = T$ outside the lens and
\begin{equation} \label{defS}
\begin{aligned}
    S &= T
    \begin{pmatrix} 1 & 0 & 0 \\ 0 & 1 & 0 \\
    - e^{n(\lambda_1-\lambda_3)} & 0 & 1 \end{pmatrix}
    \qquad \textrm{in upper part of the lens in left half-plane,} \\
    S & = T
    \begin{pmatrix} 1 & 0 & 0 \\ 0 & 1 & 0 \\
    e^{n(\lambda_1-\lambda_3)} & 0 & 1 \end{pmatrix}
    \qquad \textrm{in lower part of the lens in left half-plane,} \\
    S & = T
    \begin{pmatrix} 1 & 0 & 0 \\
    -e^{n(\lambda_1-\lambda_2)} & 1 & 0 \\ 0 & 0 & 1 \end{pmatrix}
    \qquad
    \textrm{in upper part of the lens in right half-plane,} \\
    S & = T
    \begin{pmatrix} 1 & 0 & 0 \\
    e^{n(\lambda_1-\lambda_2)} & 1 & 0 \\ 0 & 0 & 1 \end{pmatrix}
    \qquad
    \textrm{in lower part of the lens in right half-plane.}
\end{aligned}
\end{equation}

Outside the lens, the jumps for $S$ are as those for $T$, while
on $[-z_1, z_1]$ and on the upper and lower lips of the lens, the
jumps are according to the factorizations (\ref{factorization1})
and (\ref{factorization2}). The result is that $S$ satisfies the
following RH problem

\begin{itemize}
\item $S$ is analytic outside the real line, the vertical segment $[-iz_2,iz_2]$,
the curve $\Sigma$, and the upper and lower lips of the lens around $[-z_1,z_1]$.
\item $S$ satisfies the following jumps on the real line
\begin{align} \label{jumpS1}
    S_+ & = S_-
    \begin{pmatrix}
        1 & e^{n(\lambda_{2+}- \lambda_{1-})} & e^{n(\lambda_{3+}-\lambda_{1-})} \\
        0 & 1 & 0 \\
        0 & 0 & 1
        \end{pmatrix}  \qquad \textrm{on } (-\infty,-x^*] \\
    \label{jumpS2}
        S_+ & = S_-
    \begin{pmatrix}
        1 & 0 & e^{n(\lambda_{3+}-\lambda_{1-})} \\
        0 & 1 & 0 \\
        0 & 0 & 1
        \end{pmatrix} \qquad \textrm{on } (-x^*, -z_1] \\
    \label{jumpS3}
        S_+ & = S_-
        \begin{pmatrix} 0 & 0 & 1\\
        0 & 1 & 0 \\ -1 & 0 & 0 \end{pmatrix}
        \qquad \textrm{on } (-z_1,0) \\
    \label{jumpS4}
        S_+ & = S_-
    \begin{pmatrix} 0 & 1 & 0 \\
        -1 & 0 & 0 \\ 0 & 0 & 1 \end{pmatrix}
    \qquad \textrm{on } (0, z_1)  \\
      \label{jumpS5}
        S_+ & = S_-
      \begin{pmatrix}
        1 & e^{n(\lambda_2-\lambda_1)} & 0 \\
        0 & 1 & 0 \\
        0 & 0 & 1
        \end{pmatrix}
        \qquad \textrm{on } [z_1, x^*) \\
    \label{jumpS6}
        S_+ & = S_-
      \begin{pmatrix}
        1 & e^{n(\lambda_2-\lambda_1)} & e^{n(\lambda_3-\lambda_1)} \\
        0 & 1 & 0 \\
        0 & 0 & 1
        \end{pmatrix}
        \qquad \textrm{on } [x^*, \infty).
\end{align}
$S$ has the following jumps on the segment $[-iz_2,iz_2]$,
\begin{align} \label{jumpS7}
    S_+ & = S_-
    \begin{pmatrix}
        1 & 0 & 0 \\
        0 & 0 & 1 \\
        0 & -1 & e^{n(\lambda_{3+}-\lambda_{3-})}
        \end{pmatrix}  \qquad \textrm{on } (-iz_2, -iy^*) \\
        \label{jumpS8}
    S_+ & = S_-
    \begin{pmatrix}
        1 & 0 & 0 \\
        0 & 0 & 1 \\
        e^{n(\lambda_1-\lambda_{3-})} & -1 & e^{n(\lambda_{3+}-\lambda_{3-})}
        \end{pmatrix}  \qquad \textrm{on } (-iy^*,0) \\
        \label{jumpS9}
    S_+ & = S_-
    \begin{pmatrix}
        1 & 0 & 0 \\
        0 & 0 & 1 \\
        -e^{n(\lambda_1-\lambda_{3-})} & -1 & e^{n(\lambda_{3+}-\lambda_{3-})}
        \end{pmatrix}  \qquad \textrm{on } (0,iy^*) \\
    \label{jumpS10}
    S_+ & = S_-
    \begin{pmatrix}
        1 & 0 & 0 \\
        0 & 0 & 1 \\
        0 & -1 & e^{n(\lambda_{3+}-\lambda_{3-})}
        \end{pmatrix}  \qquad \textrm{on } (iy^*, iz_2).
\end{align}
The jumps on $\Sigma$ are
\begin{align}
    \label{jumpS11}
    S_+ & =  S_-
    \begin{pmatrix}
    1 & 0 & 0 \\
    0 & 1 & 0 \\
    0 & e^{n(\lambda_2-\lambda_3)} & 1
    \end{pmatrix}
    \qquad \textrm{on }  \{z \in \Sigma \mid \Re z < 0 \} \\
    \label{jumpS12}
    S_+ & =  S_-
    \begin{pmatrix}
    1 & 0 & 0 \\
    0 & 1 & e^{n(\lambda_3-\lambda_2)} \\
    0 & 0 & 1
    \end{pmatrix}
    \qquad \textrm{on } \{z \in \Sigma \mid \Re z > 0 \}.
\end{align}
Finally, on the upper and lower lips of the lens, we find jumps
\begin{align}
    \label{jumpS13}
    S_+ & = S_-
    \begin{pmatrix}
    1 & 0 & 0 \\ 0 & 1 & 0 \\ e^{n(\lambda_1-\lambda_3)} & 0 & 1
    \end{pmatrix}
    \qquad \textrm{on the lips of the lens in the left half-plane} \\
    \label{jumpS14}
    S_+ & =  S_-
    \begin{pmatrix}
    1 & 0 & 0 \\ e^{n(\lambda_1-\lambda_2)} & 1 & 0 \\ 0 & 0 & 1
    \end{pmatrix}
    \qquad \textrm{on the lips of the lens in the right half-plane.}
\end{align}
\item $S(z) = I + O(1/z)$ as $z \to \infty$.
\end{itemize}

So now we have 14 different jump matrices (\ref{jumpS1})--(\ref{jumpS14}).
As $n \to \infty$, all these jumps have limits. Most of the limits are
the identity matrix, except for the jumps on $(-z_1,z_1)$, see (\ref{jumpS3})
and (\ref{jumpS4}), and on $(-iz_2, iz_2)$, see (\ref{jumpS7})--(\ref{jumpS10}).
In the next section we will solve explicitly
the limiting model RH problem. The solution to the model problem will be
further used in the construction of parametrix away from the branch points.

\section{Parametrix away from branch points}

The model RH problem is the following.
Find $N$ such that
\begin{itemize}
\item $N : \mathbb C \setminus ([-z_1,z_1] \cup [-iz_2, iz_2]) \to \mathbb C^{3 \times 3}$ is analytic.
\item $N$ satisfies the jumps
\begin{align}
N_+ & = N_- \label{jumpN1}
    \begin{pmatrix}
    0 & 0 & 1 \\ 0 & 1 & 0 \\ -1 & 0 & 0 \end{pmatrix}
    \textrm{ on } [-z_1, 0) \\
N_+ & = N_- \label{jumpN2}
    \begin{pmatrix}
    0 & 1 & 0 \\ -1 & 0 & 0 \\ 0 & 0 & 1 \end{pmatrix}
    \textrm{ on } (0, z_1] \\
N_+ & = N_- \label{jumpN3}
    \begin{pmatrix}
    1 & 0 & 0 \\ 0 & 0 & 1 \\  0 & -1 & 0  \end{pmatrix}
    \textrm{ on } [-iz_2,iz_2].
\end{align}
\item $N(z) = I + O(1/z)$ as $z \to \infty$.
\end{itemize}

\begin{center}
 \begin{figure}[h]
 \begin{center}
  \scalebox{0.5}{\includegraphics{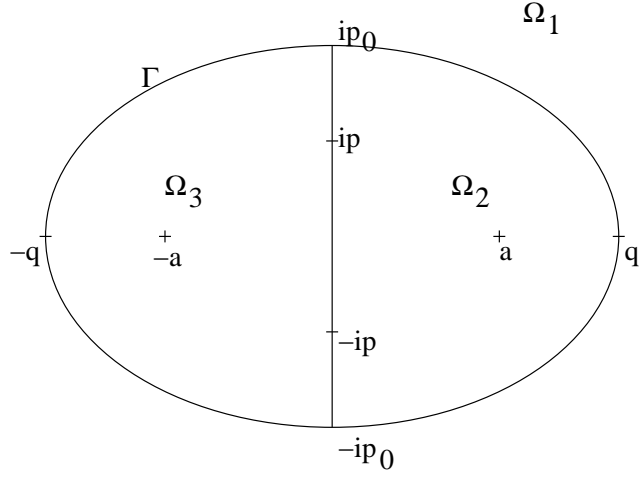}}
 \end{center}
\caption{Partition of the complex $\xi$-plane.}\label{figure6}
\end{figure}
 \end{center}

To solve the model RH problem we lift it to the Riemann surface
(\ref{cubicequation}) with the sheet structure as in Figure 1,
see also \cite{BK2}, \cite{KVAW} where the same technique was used.
Consider to that end the range of the functions $\xi_k$ on the
complex plane, $\Om_k=\xi_k(\mathbb C)$ for $k=1,2,3$.
Then $\Om_1$, $\Om_2$, $\Om_3$ give a partition of the complex
plane into three regions, see Figure \ref{figure6}. In this figure
$q$, $p$ and $p_0$ are such that
\begin{equation}
\begin{aligned}
    q & = \xi_1(z_1) = \xi_2(z_1) = -\xi_1(-z_1) = - \xi_3(-z_1), \\
    ip & = -\xi_2(iz_2) = - \xi_3(iz_2) = \xi_2(-iz_2) = \xi_3(-iz_2), \\
    ip_0 & = \xi_{1+}(0) = - \xi_{1-}(0).
\end{aligned}
\end{equation}

Let $\Ga$ be the boundary of $\Om_1$. Then we have
\begin{equation}\label{mod6}
\begin{aligned}
\xi_{1\pm}([-z_1,z_1])& =\Ga\cap\{\pm\Im z\ge 0\},
\\
\xi_{2-}([-iz_2,0])&=[ip,ip_0],\quad
\xi_{2-}([0,iz_2])=[-ip_0,-ip],
\\
\xi_{3-}([-iz_2,0])&=[ip,0],\quad
\xi_{3-}([0,iz_2])=[0,-ip],
\end{aligned}
\end{equation}
and $\xi_{2\pm}(iy)=\xi_{3\mp}(iy)$ for $-z_2\le y\le z_2$.
According to our agreement,
on the interval $-iz_2\le y\le iz_2$ the minus side is on the right.

We are looking for a solution $N$ in the following form:
\begin{equation}\label{mod7}
N(z)=
\begin{pmatrix}
N_1(\xi_1(z)) & N_1(\xi_2(z)) & N_1(\xi_3(z)) \\
N_2(\xi_1(z)) & N_2(\xi_2(z)) & N_2(\xi_3(z)) \\
N_3(\xi_1(z)) & N_3(\xi_2(z)) & N_3(\xi_3(z))
\end{pmatrix},
\end{equation}
where $N_1(\xi)$, $N_2(\xi)$, $N_3(\xi)$ are three scalar analytic functions
on $\mathbb C \setminus (\Ga \cup [-ip_0,ip_0])$. To satisfy the jump conditions
on $N(z)$ we need the following jump relations for
$N_j(\xi)$, $j=1,2,3$:
\begin{equation}\label{mod9}
\begin{aligned}
N_{j+}(\xi)&=N_{j-}(\xi),\quad \xi\in\left(\Ga\cap\{\Im z\le 0\}\right)
\cup[-ip_0,-ip]\cup[ip,ip_0], \\
N_{j+}(\xi)&=-N_{j-}(\xi),\quad \xi\in\left(\Ga\cap\{\Im z\ge 0\}\right)
\cup [-ip,ip].
\end{aligned}
\end{equation}
So the $N_j$'s are actually analytic across the curve $\Gamma$ in the lower
half-plane and on the segments $[ip,ip_0]$ and $[-ip_0, -ip]$. What remains
are the curve $\Gamma$ in the upper half-plane and the segment $[-ip,ip]$,
where the functions change sign.
Since $\xi_1(\infty)=\infty$, $\xi_2(\infty)=a$, $\xi_3(\infty)=-a$,
then to satisfy $N(\infty)=I$ we require
\begin{equation}\label{mod10}
\begin{aligned}
N_1(\infty)&=1,\quad N_1(a)=0,\quad N_1(-a)=0;\\
N_2(\infty)&=0,\quad N_2(a)=1,\quad N_2(-a)=0;\\
N_3(\infty)&=0,\quad N_3(a)=0,\quad N_3(-a)=1.
\end{aligned}
\end{equation}
Thus, we obtain three scalar RH problems on $N_1,N_2,N_3$.
Equations (\ref{mod9})--(\ref{mod10}) have the following solution:
\begin{equation}\label{mod11}
N_1(\xi)=\frac{\xi^2-a^2}
{\sqrt{(\xi^2+p^2)(\xi^2-q^2)}},\quad
N_{2,3}(\xi)=c_{2,3}\frac{\xi \pm a}
{\sqrt{(\xi^2+p^2)(\xi^2-q^2)}},
\end{equation}
with cuts at $\Ga\cap\{\Im \xi\ge 0\}$ and  $[-ip,ip]$.
The constants $c_{2,3}$ are determined
by the equations $N_{2,3}(\pm a)=1$. We have that
\begin{equation}\label{mod12}
(\xi^2+p^2)(\xi^2-q^2)
=\xi^4-(1+2a^2)\xi^2+(a^2-1)a^2 \equiv R(\xi;a),
\end{equation}
and as in Section 6 of \cite{BK2}, we obtain
$ c_2 = c_3 =-\frac{i}{\sqrt 2}$.
Thus, the solution to the model RH problem is given by
\begin{equation}\label{mod17}
N(z)=
\begin{pmatrix}
\frac{\xi_1^2(z)-a^2}{\sqrt{R(\xi_1(z);a)}}
& \frac{\xi_2^2(z)-a^2}{\sqrt{R(\xi_2(z);a)}}
& \frac{\xi_3^2(z)-a^2}{\sqrt{R(\xi_3(z);a)}} \\
-i\frac{\xi_1(z)+a}{\sqrt{2R(\xi_1(z);a)}}
& -i\frac{\xi_2(z)+a}{\sqrt{2R(\xi_2(z);a)}}
& -i\frac{\xi_3(z)+a}{\sqrt{2R(\xi_3(z);a)}} \\
-i\frac{\xi_1(z)-a}{\sqrt{2R(\xi_1(z);a)}}
& -i\frac{\xi_2(z)-a}{\sqrt{2R(\xi_2(z);a)}}
& -i\frac{\xi_3(z)-a}{\sqrt{2R(\xi_3(z);a)}}
\end{pmatrix},
\end{equation}
with cuts on $[-z_1,z_1]$ and $[-iz_2,iz_2]$.

\section{Local parametrices}
Near the branch points $N$ will not be a good approximation to $S$.
We need a local analysis near each of the branch points.
In a small circle around each of the branch points, the parametrix $P$
should have the same jumps as $S$, and on the boundary of the
circle $P$ should match with $N$ in the sense that
\begin{equation} \label{matching}
    P(z) = N(z) \left(I + O(1/n)\right)
\end{equation}
uniformly for $z$ on the boundary of the circle.

The construction of $P$ near the real branch points $\pm z_1$ makes use
of Airy functions and it is the same as the one given
in \cite[Section 7]{BK2} for the case $a > 1$.
The parametrix near the imaginary branch points $\pm i z_2$
is also constructed with Airy functions. We give the construction
near $i z_2$. We want an analytic $P$ in a neigborhood of $i z_2$
with jumps
\begin{equation} \label{jumpsP}
\begin{aligned}
    P_+ & = P_-
    \begin{pmatrix}
    1 & 0 & 0 \\ 0 & 1 & 0 \\
    0 & e^{n(\lambda_2-\lambda_3)} & 1
    \end{pmatrix}
    \qquad \textrm{on  left contour} \\
    P_+ & = P_-
    \begin{pmatrix}
    1 & 0 & 0 \\ 0 & 1 & e^{n(\lambda_3-\lambda_2)} \\
    0 & 0 & 1 \end{pmatrix}
    \qquad \textrm{on right contour} \\
    P_+ & = P_-
    \begin{pmatrix}
    1 & 0 & 0 \\ 0 & 0 & 1 \\
    0 & -1 & e^{n(\lambda_{3+}-\lambda_{3-})}
    \end{pmatrix}
    \qquad \textrm{on vertical part.}
\end{aligned}
\end{equation}
In addition we need the matching condition (\ref{matching}).
Except for the matching condition (\ref{matching}),
the problem is a $2\times 2$ problem.

Let us consider  $\lambda_2 - \lambda_3$ near the branch point $iz_2$.
We know that $(\lambda_2 - \lambda_3)(iz_2) = 0$, see (\ref{jumpslambda})
and since $\xi_2 - \xi_3$ has square root behavior
at $iz_2$ it follows that
\[  (\lambda_2 - \lambda_3)(z)
    = \int_{iz_2}^z (\xi_2(s) - \xi_3(s)) ds =
    (z-iz_2)^{3/2} h(z)
\]
with an analytic function $h$ with $h(iz_2) \neq 0$.
So we can take a $2/3$-power and obtain a conformal map.
To be precise, we note that
\[ \arg ((\lambda_2 - \lambda_3)(iy))  = \pi/2, \qquad
    \textrm{ for } y > z_2, \]
and so we define
\begin{equation} \label{deffz}
    f(z) = \left[\frac{3}{4} (\lambda_2-\lambda_3)(z) \right]^{2/3}
\end{equation}
such that
\[ \arg f(z) = \pi/3, \qquad \textrm{ for } z = iy, \, y > z_2. \]
Then $s = f(z)$ is a conformal map, which maps $[0,iz_2]$ into the
ray $\arg s = - \frac{2\pi}{3}$, and which maps
the parts of $\Sigma$ near $iz_2$ in the right and left half-planes
into the rays $\arg s =0$ and $\arg s = \frac{2\pi}{3}$,
respectively.
[Recall that $\lambda_2 - \lambda_3$ is real on these
contours, see (\ref{constrSigma}).]

We choose $P$ of the form
\begin{equation} \label{ansatzPz}
    P(z) = E(z) \Phi\left(n^{2/3} f(z)\right)
    \begin{pmatrix} 1 & 0 & 0 \\ 0 & e^{\frac{1}{2} n (\lambda_2-\lambda_3)} & 0 \\
    0 & 0 & e^{-\frac{1}{2}n (\lambda_2-\lambda_3)}
    \end{pmatrix}
\end{equation}
where $E$ is analytic. In order to satisfy the jump conditions (\ref{jumpsP})
we want that $\Phi$ is defined and analytic in the complex $s$-plane
cut along the three rays $\arg s = k \frac{2 \pi i}{3}$, $k = -1, 0,1$,
and there it has jumps
\begin{equation} \label{jumpsPhi}
\begin{aligned}
    \Phi_+ & = \Phi_- \begin{pmatrix}
        1 & 0 & 0 \\ 0 & 1 & 0 \\ 0 & 1 & 1 \end{pmatrix}
        \qquad \textrm{for } \arg s = 2\pi/3, \\
    \Phi_+ & = \Phi_- \begin{pmatrix}
        1 & 0 & 0 \\ 0 & 0 & 1 \\ 0  & -1 & 1 \end{pmatrix}
        \qquad \textrm{for } \arg s = -2\pi/3, \\
    \Phi_+ & = \Phi_- \begin{pmatrix}
        1 & 0 & 0 \\ 0 & 1 & 1 \\ 0 & 0 & 1 \end{pmatrix}
        \qquad \textrm{for } \arg s =0.
\end{aligned}
\end{equation}

Put $y_0(s) = \Ai(s)$, $y_1(s) = \omega \Ai(\omega s)$,
$y_2(s) = \omega^2 \Ai(\omega^2 s)$ with $\omega = 2\pi/3$
and $\Ai$ the standard Airy function.
Then we take $\Phi$ as
\begin{equation} \label{defPhi}
\begin{aligned}
    \Phi&= \begin{pmatrix}
    1 & 0 & 0 \\ 0 & y_0 & - y_2 \\ 0 & y_0' & -y_2'
    \end{pmatrix}
    \qquad \textrm{for } 0 < \arg s < 2\pi/3, \\
    \Phi&= \begin{pmatrix}
    1 & 0 & 0 \\ 0 & y_0 & y_1 \\ 0 & y_0' & y_1'
    \end{pmatrix}
    \qquad \textrm{for } -2\pi/3 < \arg s < 0, \\
    \Phi&= \begin{pmatrix}
    1 & 0 & 0 \\ 0 & - y_1 & - y_2 \\ 0 & -y_1' & -y_2'
    \end{pmatrix}
    \qquad \textrm{for } 2\pi/3 < \arg s < 4\pi/3.
\end{aligned}
\end{equation}
This $\Phi$ satisfies the jumps (\ref{jumpsPhi}). In order to achieve
the matching (\ref{matching}) we define the prefactor $E$ as
\begin{equation} \label{defE}
    E =  N L^{-1}
\end{equation}
with
\begin{equation} \label{defL}
    L = \frac{1}{2 \sqrt{\pi}} \begin{pmatrix} 1 & 0 & 0 \\
        0 & n^{-1/6} f^{-1/4} & 0 \\
        0 & 0 & n^{1/6} f^{1/4} \end{pmatrix}
         \begin{pmatrix} 1 & 0   & 0 \\
        0 & 1 & i \\ 0 & -1 & i \end{pmatrix}
\end{equation}
where $f^{1/4}$ has a branch cut along the vertical segment
$[0, iz_2]$ and it is real and positive where $f$ is
real and positive.
The matching condition (\ref{matching}) now follows from the
asymptotics of the Airy function and its derivative
\[
\begin{aligned}
    \Ai(s) & = \frac{1}{2\sqrt{\pi} } s^{-1/4}e^{-\frac{2}{3} s^{3/2}}
    \left(1+ O(s^{-3/2}) \right), \\
    \Ai'(s) & = -\frac{1}{2 \sqrt{\pi}} s^{1/4} e^{-\frac{2}{3} s^{3/2}}
    \left(1+O(s^{-3/2}) \right),
    \end{aligned}
    \]
as $s \to \infty$, $|\arg s| < \pi$.
On the cut we have $f^{1/4}_+ = i f^{1/4}_-$. Then (\ref{defL})
gives
\[ L_+ = L_- \begin{pmatrix} 1 & 0 & 0 \\ 0 & 0 & 1 \\ 0 & -1 & 0
    \end{pmatrix}, \]
which is the same jump as satisfied by $N$, see (\ref{jumpN3}).
This implies that $E = N L^{-1}$ is analytic in a punctured neighborhood
of $iz_2$. Since the entries of $N$ and $L$ have at most fourth-root
singularities, the isolated singularity is removable, and
$E$ is analytic. It follows that $P$ defined by (\ref{ansatzPz})
does indeed satisfy the jumps (\ref{jumpsP}) and the
matching condition (\ref{matching}).

A similar construction gives the parametrix in the
neighborhood of $-iz_2$.

\section{Fourth transformation $S \mapsto R$}
Having constructed $N$ and $P$, we define the final transformation by
\begin{equation}
\begin{aligned}
R(z) & = S(z) N(z)^{-1} \qquad \textrm{away from the branch points,} \\
R(z) & = S(z) P(z)^{-1} \qquad \textrm{near the branch points}.
\end{aligned}
\end{equation}
Since jumps of $S$ and $N$ coincide on the interval $(-z_1, z_1)$
and the jumps of $S$ and $P$ coincide inside the disks around the
branch points, we obtain that $R$ is analytic outside a system of
contours as shown in Figure 7.

  \begin{figure}[h] \label{figure7}
  \centerline{
  \includegraphics[width=16cm,height=8cm]{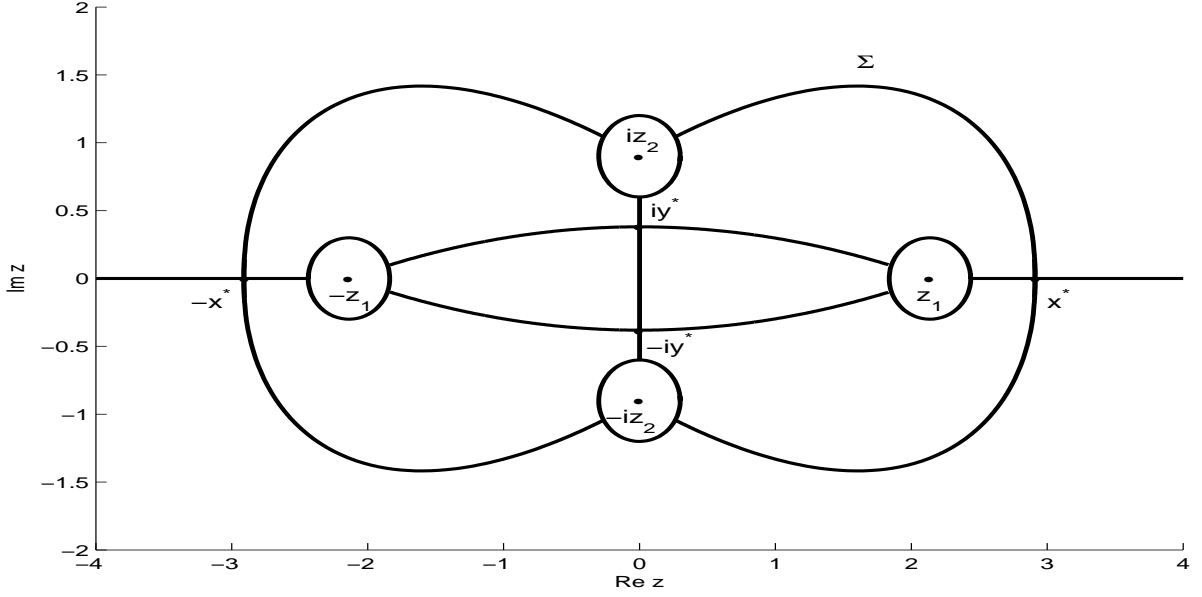}}
        \caption{$R$ has jumps on this system of contours.}
    \end{figure}

On the circles around the branch points there is a jump
\begin{equation} \label{jumpsR1}
    R_+ = R_- (I + O(1/n)),
 \end{equation}
which follows from the matching condition (\ref{matching}).
On the remaining contours, the jump is
\begin{equation} \label{jumpsR2}
    R_+ = R_- (I + O(e^{-cn}))
\end{equation}
for some $c > 0$.
Since we also have the asymptotic condition $R(z) = I + O(1/z)$
as $z \to \infty$, we may conclude as in \cite[Section 8]{BK1} that
\begin{equation} \label{Rasymp}
    R(z) = I + O\left(\frac{1}{n(|z|+1)} \right) \qquad \mbox{ as } n \to \infty,
\end{equation}
uniformly for $z \in \mathbb C$, see also \cite{Deift,DKMVZ1,DKMVZ2,Kui}.

\section{Proof of Theorems \ref{maintheo1} and \ref{maintheo2}}

We follow the expression for the kernel $K_n$ as we make
the transformations $Y \mapsto U \mapsto T \mapsto S$.
From (\ref{kernelK}) and the transformation (\ref{defU})
it follows that $K_n$ has the following expression in terms of $U$,
for any $x, y \in \mathbb R$,
\begin{equation} \label{kernelK2}
    K_n(x,y) = \frac{e^{\frac{1}{4}n(x^2-y^2)}}{2\pi i (x-y)}
    \begin{pmatrix} 0 & e^{n\lambda_{2+}(y)} & e^{n\lambda_{3+}(y)} \end{pmatrix}
    U_+^{-1}(y) U_+(x)
    \begin{pmatrix} e^{-n\lambda_{1+}(x)} \\ 0 \\ 0 \end{pmatrix}.
\end{equation}
Then from (\ref{defT}) we obtain for  $y \geq 0$ inside the contour
$\Sigma$, and for any $x \in \mathbb R$,
\begin{equation} \label{kernelK3}
K_n(x,y)  = \frac{e^{\frac{1}{4}n(x^2-y^2)}}{2\pi i (x-y)}
    \begin{pmatrix} 0 & e^{n\lambda_{2+}(y)} & 0 \end{pmatrix}
    T_+^{-1}(y) T_+(x)
    \begin{pmatrix} e^{-n\lambda_{1+}(x)} \\ 0 \\ 0 \end{pmatrix},
\end{equation}
and from (\ref{defS}), we have when $x,y \in [0,z_1)$,
\begin{equation} \label{kernelK4}
K_n(x,y) = \frac{e^{\frac{1}{4}n(x^2-y^2)}}{2\pi i (x-y)}
    \begin{pmatrix} -e^{n\lambda_{1+}(y)}  & e^{n\lambda_{2+}(y)} & 0 \end{pmatrix}
    S_+^{-1}(y) S_+(x)
    \begin{pmatrix} e^{-n\lambda_{1+}(x)} \\ e^{-n\lambda_{2+}(x)} \\ 0 \end{pmatrix}.
\end{equation}
Since $\lambda_{1+}$ and $\lambda_{2+}$ are each others complex conjugates on
$[0,z_1)$, we can rewrite (\ref{kernelK4}) for $x, y \in [0,z_1)$ as
\begin{equation} \label{kernelK5}
K_n(x,y) = \frac{e^{n(h(y)-h(x))}}{2\pi i(x-y)}
    \begin{pmatrix} - e^{n i \im \lambda_{1+}(y)} & e^{-n i \im \lambda_{1+}(y)} & 0 \end{pmatrix}
    S_+^{-1}(y) S_+(x)
    \begin{pmatrix} e^{-n i \im \lambda_{1+}(x)} \\ e^{n i \im \lambda_{1+}(x)} \\ 0 \end{pmatrix},
\end{equation}
where
\begin{equation} \label{defh}
    h(x) = \Re \lambda_{1+}(x) - \frac{1}{4}x^2.
\end{equation}
Note that (\ref{kernelK5}) is exactly the same as equation (5.14) in \cite{BK2}.
Therefore we can almost literally follow the proofs in Section 9 of \cite{BK2}
to complete the proof of Theorem \ref{maintheo1} and \ref{maintheo2}.

Indeed as in \cite{BK2} the limiting mean density (\ref{rho1})
follows from (\ref{kernelK5}) and (\ref{Rasymp}) in case $x > 0$,
where
\begin{equation} \label{rho2}
    \rho(x) = \frac{1}{\pi} \Im \xi_{1+}(x), \qquad x \in \mathbb R.
\end{equation}
The case $x < 0$ follows in the same way and also by symmetry.
Recalling that the choice of the cut $[-iz_2,iz_2]$ was arbitrary
as remarked at the end of section 2, we note that we might as well
have done the asymptotic analysis on a contour that does not pass
through $0$, so that we obtain (\ref{rho1}) for $x =0$ as well.
The statement in part (a) on the behavior of $\rho$ follows immediately
from (\ref{rho2}) and the properties of $\xi_1$ as the inverse
mapping of (\ref{mapping}).

This completes the proof of Theorem \ref{maintheo1}.

\medskip

The proof of part (a) of Theorem \ref{maintheo2} for the case $x_0 > 0$
follows from (\ref{kernelK5}) and (\ref{Rasymp}) exactly
as in Section 9 of \cite{BK2}. The case $x_0 < 0$
follows by symmetry, and the case $x_0 = 0$ follows as well, since
we might have done the asymptotic analysis on a cut different
from $[-iz_2, iz_2]$, as just noted above.
The proof of part (b) follows as in \cite{BK2} as well.
 Note however that the proof of part (b) relies on
the local parametrix at the branch point $z_1$, which we have not
specified explicitly in Section 7. However, the formulas
are the same as those in \cite{BK2} and the proof can be copied.
This completes the proof of Theorem \ref{maintheo2}.

\end{document}